\documentstyle[pramana,epsf,floats,amssymb]{ias}
\def\Ob{\Omega_b}

\def\Ok{\Omega_k}
\def\Ol{\Omega_\Lambda}
\def\Om{\Omega_m}
\def\Od{\Omega_d}

\def\ob{\omega_b}

\def\od{\omega_d}

\def\om{\omega_{\rm m}}

\def\fn{f_\nu}
\def\ns{{n_s}}
\def\nt{{n_t}}

\def\Ot{\Omega_{\rm tot}}
\def\As{A_s}
\def\At{A_t}

\def\zion{z_{\rm ion}}

\def\Mnu{M_\nu}

\def\plotone#1{\centering \leavevmode
\epsfxsize= 0.8\columnwidth \epsfbox{#1}}

\def\plottwo#1#2{\leavevmode
\epsfxsize=.45\columnwidth \epsfbox{#1} \centering
\epsfxsize=.45\columnwidth \epsfbox{#2}}

\def\lsim{\mathrel{\rlap{\lower4pt\hbox{\hskip1pt$\sim$}}
    \raise1pt\hbox{$<$}}}                % less than or approx. symbol
\def\gsim{\mathrel{\rlap{\lower4pt\hbox{\hskip1pt$\sim$}}
    \raise1pt\hbox{$>$}}}                % greater than or approx. symbol

\begin{document}

\mark{{Observational cosmology}{Jeremiah P Ostriker 
and  Tarun Souradeep }} 

\title{The current status of observational cosmology }

\author{JEREMIAH P OSTRIKER$^1$  and TARUN SOURADEEP$^2$}

\address{$^1$Department of Astrophysical Sciences, Peyton
Hall, Princeton University, \\ Princeton, NJ 08544, USA\\
$^2$Inter-University Centre for Astronomy and
Astrophysics,\\ Post Bag 4, Ganeshkhind,  Pune 411 007, India\\
E-mail: jpo@astro.princeton.edu,tarun@iucaa.ernet.in}

\keywords{cosmology, observations}

 \pac{98.8Es}

\abstract{Observational cosmology has indeed made very rapid progress
in recent years. The ability to quantify the universe has largely
improved due to observational constraints coming from structure
formation. The transition to precision cosmology has been spearheaded
by measurements of the anisotropy in the cosmic microwave background
(CMB) over the past decade. Observations of the large scale structure
in the distribution of galaxies, high red-shift supernova, have
provided the required complementary information.  We review the
current status of cosmological parameter estimates from joint analysis
of CMB anisotropy and large scale structure (LSS) data. We also sound
a note of caution on overstating the successes achieved thus far.}

\maketitle

\section{Introduction} 

Recent developments in cosmology have been largely driven by huge
improvement in quality, quantity and the scope of cosmological
observations.  The measurement of temperature anisotropy in the cosmic
microwave background (CMB) has been arguably the most influential of
these recent observational success stories.  A glorious decade of CMB
anisotropy measurements has been topped off by the data from the
Wilkinson Microwave Anisotropy Probe (WMAP) of NASA. Observational
success has set off an intense interplay between theory and
observations. While the observations have constrained theoretical
scenarios and models more precisely, some of these observations have
thrown up new challenges to theoretical understanding and others that
have brought issues from the realm of theoretical speculation to
observational verification. The results the WMAP mission on CMB
anisotropy~\cite{ben_wmap03} and the power spectrum of density
perturbations from the 2-degree field (2dF) survey of galaxies
\cite{2df} and the Sloan Digital Sky Survey (SDSS) of
galaxies\cite{sdss}, have allowed very precise estimation of
cosmological parameters~\cite{sper_wmap03,maxsdss,sel04}.

These results have been widely heralded as the dawn of precision
cosmology.  To a casual science observer, the unprecedented precision
in determining the parameters of the standard cosmological model often
conveys the false impression that we actually know and understand the
components that make up the universe, we know its primeval history
(including the theoretical scenarios of baryogenesis and inflation) and 
evolution including growth of large scale structures. The reality is
that our understanding of some of the components is limited to very
rudimentary characterization, e.g., in terms of their cosmic energy
density, velocity dispersion,   equation of state etc. Besides the
obvious need for direct detection, we are not quite in a position 
even to rule out equally viable non-standard alternatives.

The cosmological model can be broadly split into two distinct aspects:
the nature and dynamics of the homogeneous background and, the origin
and evolution of perturbations leading to the large scale structure in
the distribution of matter in the universe. It is certainly fair to
say that the present edifice of the standard cosmological models is
robust.  A set of foundation and pillars of cosmology have emerged and
are each supported by a number of distinct observations:
\begin{itemize}
\itemsep-2pt
\item{}Homogeneous, isotropic cosmology, expanding from a hot initial
phase due to gravitational dynamics of the Friedman equations derived
from the laws of general relativity.

\item{} The basic constituents of the universe are baryons, photons,
neutrinos, dark matter and dark energy (cosmological constant/vacuum
energy).
 
\item{} The homogeneous spatial sections of space-time are nearly
geometrically flat (Euclidean).
 
\item{} Generation of primordial perturbations in an inflationary
epoch.  The primordial density perturbations are adiabatic with a
nearly scale invariant power spectrum. Imprint of these perturbations
on CMB anisotropy indicates correlation on a scale larger than the
causal horizon.  Polarization of the CMB anisotropy provides even
stronger support for adiabatic initial conditions and the apparently
`acausal' correlation in the primordial perturbations.

\item{} Evolution of density perturbations under gravitational
instability has produced the large scale structure in the distribution
of matter.
\end{itemize}

The past few years have seen the emergence of a `concordant'
cosmological model that is consistent with observational
constraints from the background evolution of the universe as well 
as with those from the formation of large scale structures.

The emergent concordance cosmological model does face challenges from
future observations. For example, the detection of the inflationary
gravity wave in B-mode of CMB polarization would be needed to clinch
the case for inflation.  The current observations have also revealed
potential cracks in the cosmological model which need to be resolved
through improved theoretical computations and improved future
observations. An example is the inability to recover the profile and
number density of cusps in the halos with the standard
(collision-less) cold dark matter.

\section{Observations of the cosmological model }

The evolution of the universe is an initial value problem in general
relativity that governs Einstein's theory of gravitation. Under the
assumptions of large scale homogeneity and isotropy of space (spatial
sections in a foliation of space-time), the dynamics of the spatial
sections reduces to the time evolution of the scale factor $a(t)$ of
the spatial section. Further, the components (species) of matter are
assumed to be non-interacting (on cosmological scales), ideal,
hydrodynamic fluids, specified by their energy/mass density $\rho$ and
the pressure $p$ (equivalently, by the equation of state $w$ where
$p=w\rho$). The density of any component $i$ then evolves in the
expanding universe as $\rho_i=\rho_{\rm 0i} a^{-3(1+w_i)}$. The simple
Friedman equation

\begin{figure}[t!]
 \plottwo{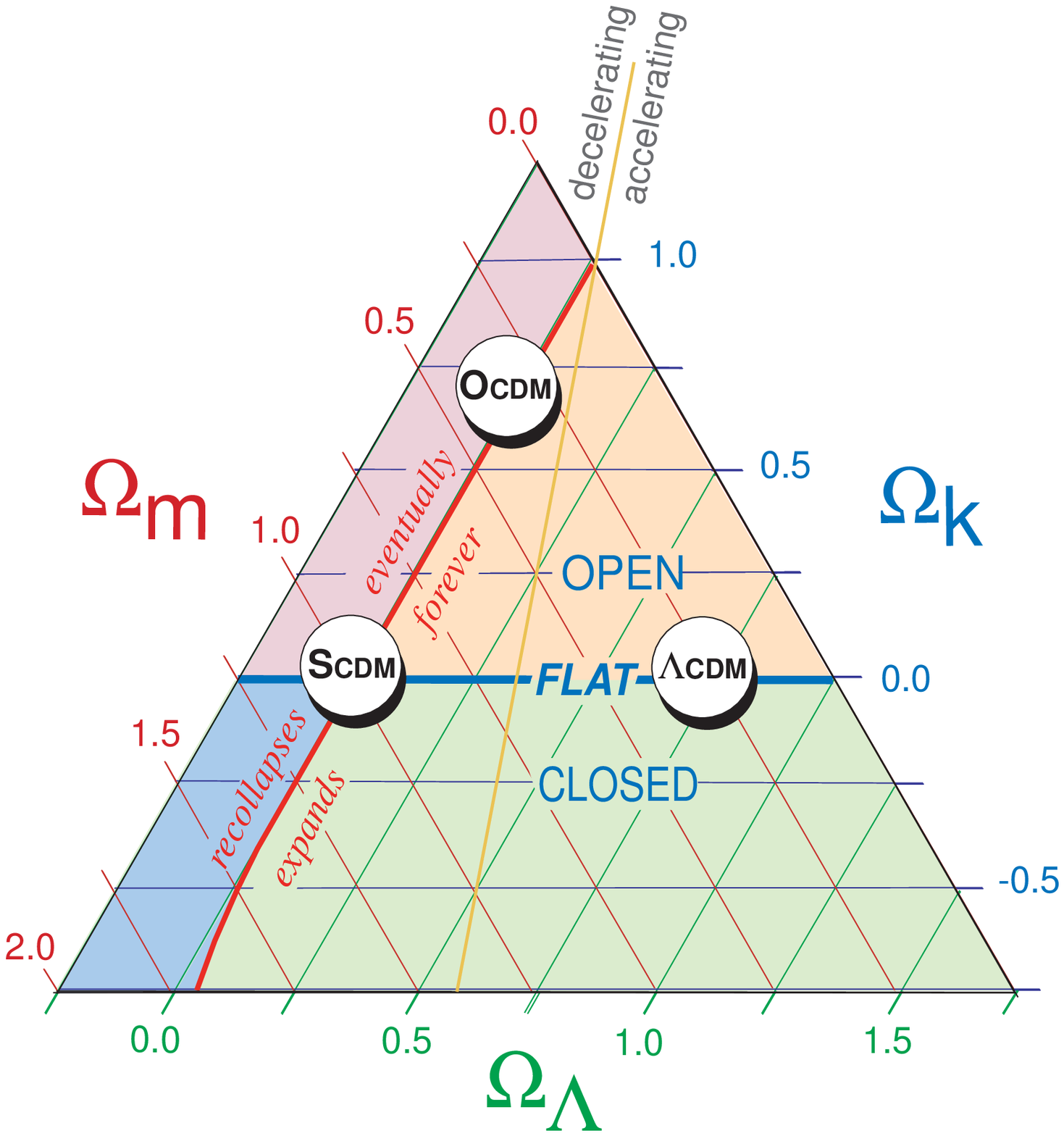}{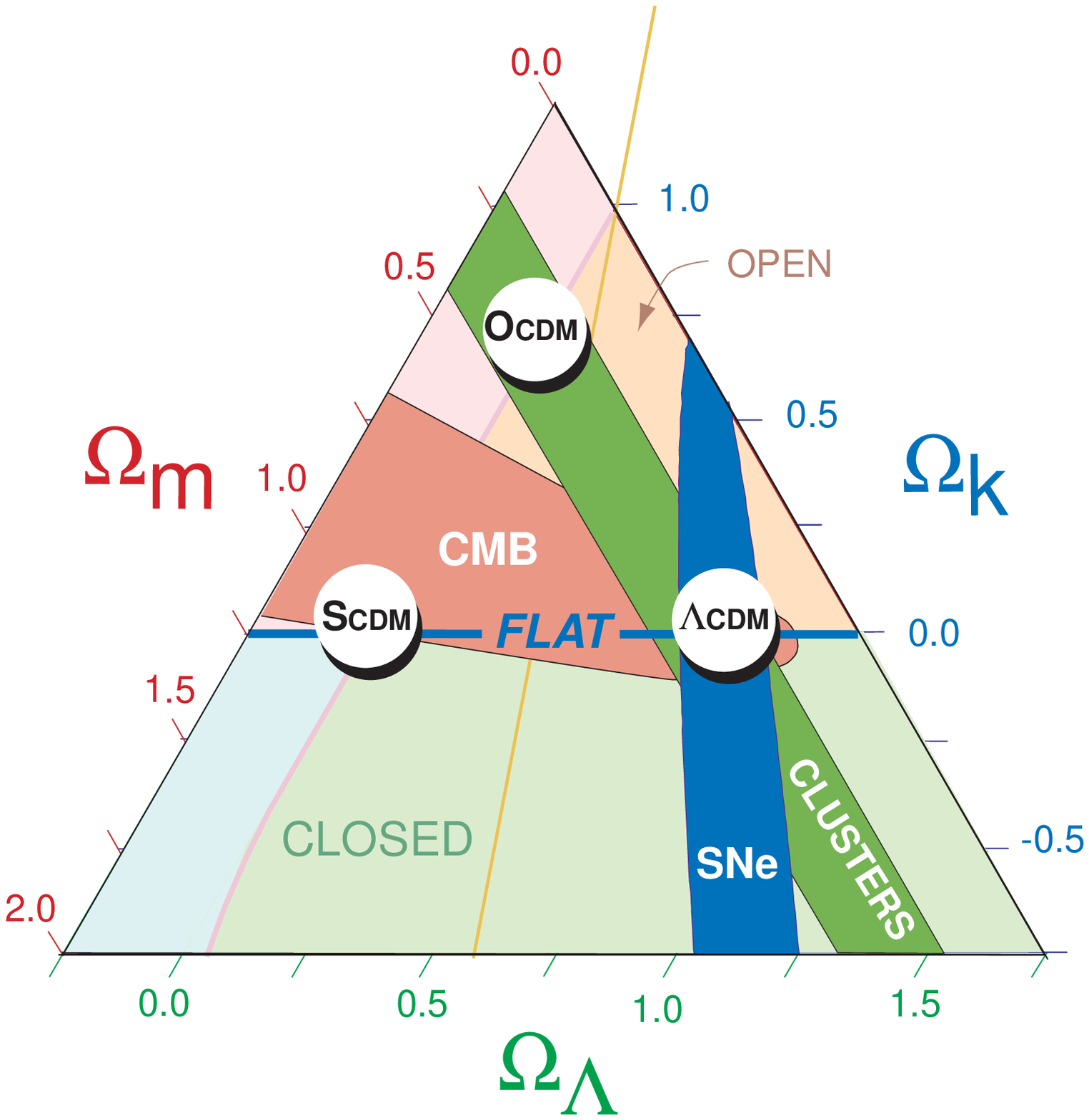}
\vspace{0.8em}

\caption{{\bf The Cosmic Triangle} represents the three key
cosmological parameters -- $\Omega_m$, $\Omega_{\Lambda}$, and
$\Omega_k$ -- where each point in the triangle satisfies the sum rule
$\Omega_m + \Omega_{\Lambda} + \Omega_k = 1$.  The blue horizontal
line (marked Flat) corresponds to a flat universe ($\Omega_m +
\Omega_{\Lambda} = 1$), separating an open universe from a closed one.
The red line, nearly along the $\Lambda=0$ line, separates a universe
that will expand forever (approximately $\Omega_{\Lambda}>0$) from one
that will eventually recollapse (approximately
$\Omega_{\Lambda}<0$). And the yellow, nearly vertical line separates
a universe with an expansion rate that is currently decelerating from
one that is accelerating.  The location of three key models are
highlighted: (flat) standard cold-dark-matter ({\sc Scdm}) ; flat
($\Lambda${\sc cdm}); and Open CDM ({\sc Ocdm}).  The right panel
schematically shows the complementarity of constraints ($\sim 1\sigma$)
from LSS, High-z SN Ia, and CMB anisotropy are shown by the three
color bands.The three independent bands intersect at the emergent
concordance model -- flat model with $\Omega_m\sim 0.3$ and
$\Omega_{\Lambda}= 0.7$. }
\label{cosmictriangle}
\end{figure}

\begin{equation}
H^2(t) \equiv \left(\frac{\dot a}{a}\right)^2 = \frac{8\pi
G}{3}~\left[ \rho_{\rm 0 m} a^{-3}+ \rho_{\rm 0 r} a^{-4} +
\rho_\Lambda + \rho_{\rm 0 K} a^{-2}\right]
\label{friedeq}
\end{equation}
that arises from the Einstein equations relates the Hubble parameter
$H(t)$ that measures the expansion rate of the universe to the present
matter density in the universe. Dividing by $H^2$ on both sides leads
to a simple sum rule
\begin{equation}
\Omega_{\rm m} +\Omega_{\rm r} + \Omega_\Lambda +
\Omega_{\rm K} = 1\,,
\label{fried}
\end{equation}
where we use the conventional dimensionless density parameter
$\Omega_i= \rho_i/\rho_{\rm c}$ in terms of the critical density
$\rho_{\rm c}=3H^2/8\pi G$.  The key components of the universe are
relativistic matter (eg., radiation) $\Omega_{\rm r}$, pressure-less
gravitating matter $\Omega_{\rm m}$, cosmological vacuum (dark)
energy, $\Omega_\Lambda$. The departure of the total matter density
parameter from unity contributes to the curvature of the space and
can, hence, be represented by an effective curvature energy density,
$\Omega_K$ that determines the effect of curvature on the expansion of
the universe. The relativistic matter density is almost entirely
dominated by the cosmic microwave background (CMB) and the relic
background of massless neutrinos.  The isentropic expansion dictated
by the Friedman equations implies that although $\Omega_{\rm r}$ makes
a negligible contribution at present given by the temperature of the
CMB, at an early epoch the universe was dominated by relativistic
matter density. The pressure-less matter density $\Omega_{\rm m} =
\Omega_{\rm B}+\Omega_{\rm cdm}+\Omega_\nu$ minimally consists of
three distinct components, the baryonic matter, cold dark matter, and
a minor contribution from massive neutrino species.  The dark energy
density could well be an exotic, `non-clustering' matter with a
variable equation of state, $w(z) < -1/3$. In this article, we limit
our attention to the simplest case of a cosmological constant with a
constant equation of state, $w=-1$.

The present state of the universe in terms the three dominant
components can neatly be summarized on the `cosmic triangle' shown in
figure 1~\cite{bah99}. The three axes address fundamental issues
regarding the background cosmology -- does space have positive,
negative or zero curvature, whether the expansion is accelerating or
decelerating, and, the issue of the non-relativistic matter budgets?
The current cosmological observations have definitively determined the
present universe to be located in the $\Lambda$-{\sc cdm} region and
far removed from the past favorites, the canonical standard cold dark
matter ({\sc Scdm}) and open cold dark matter ({\sc Ocdm}) models with
high statistical significance.

Some of the cosmological parameters are well-estimated by observations
that probe the background evolution of the universe.  The latest
constraint on the baryon density, $\Omega_{\rm B} h^2=0.022\pm 0.002$
obtained by matching the predicted abundances of light elements from
Big Bang nucleosynthesis with observations~\cite{bbn} is consistent
with that recently obtained from considerations of structure
formation~\cite{sper_wmap03,maxsdss,sel04}.

The energy density of the cosmological constant (or, more broadly
quintessence) can be inferred from the measurement of luminosity
distance as a function of red-shift using high red-shift supernova SN Ia
as standard candles.  The recent results using supernova indicate that
$\Omega_\Lambda - 1.4 \Omega_{\rm m}=0.35 \pm 0.14$~\cite{ton_SN03}.  For a
flat universe (indicated by CMB+LSS data), $\Omega_{\rm m}
=0.29\pm0.05$ and 
the equation of state $w=-1.02\pm 0.2$ is consistent with a
cosmological constant. Further, the analysis including new SN Ia
observations from HST concludes that allowing for a variable equation
of state, $w(z)$, there is no evidence for any rapid
variation~\cite{reissSN04}. As mentioned below, this is consistent with
the constraints from the CMB anisotropy and large scale structure
observations and combined constraints are remarkably tight around the
cosmological constant.

Measuring the expansion rate of the universe, $H_0=100 h$ km
s$^{-1}$/{Mpc} was a key project of the Hubble space telescope
mission. The current estimates are $h = 0.72\pm
0.07$~\cite{hst_h0}. The high-$z$ SN Ia results also constrain the
combination $H_0 t_0 = 0.96 \pm 0.04$ implying an age $t_0 = 13.6 \pm
1.5 \,\,\,\rm Gyr$ when combined with the HST determination of $H_0$.  The
expansion rate and age estimates are again consistent with, and
considerably improved by including structure formation consideration
as discussed below.

\section{Structure formation in the universe}

The `standard' model of cosmology must not only explain the dynamics
of homogeneous background universe, but also (eventually)
satisfactorily describe the perturbed universe -- the generation,
evolution and finally, the formation of large scale structures in the
universe. It is fair to say that much of the recent progress in
cosmology has come from the interplay between refinements of the
theories of structure formation and the improvement of the
corresponding observations.

The CMB anisotropies are the imprints of the perturbed universe in the
radiation. On the large angular scales, the CMB anisotropy directly probes
the primordial power spectrum on scales enormously larger than the
`causal horizon' at the epoch of last scattering at a red-shift, $z\sim
1100$. On smaller angular scales, the CMB temperature fluctuations
probe the physics of the coupled baryon--photon fluid through the
imprint of the acoustic oscillations in the ionized plasma sourced by
the same primordial fluctuations. The physics of CMB anisotropy is
well-understood, the predictions (of the linear primary anisotropy)
and their connection to observables are
unambiguous~\cite{bon_LH96,hu_dod02}. The remarkable success of the
measurements of the CMB angular power spectrum $C_l$ over the past
decade leading up to the recent WMAP results is covered elsewhere in
these proceedings (see  p. --- of {\it Pramana -- J. Phys.}
{\bf 64(4)}, (2004)).

\begin{figure}[t!]
\plotone{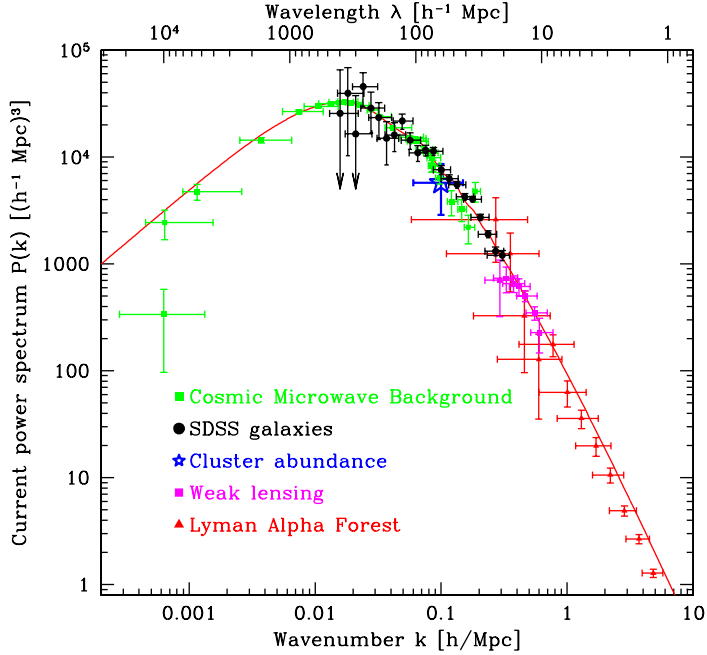}
\caption{ Figure taken from [2] shows that the current
estimate of the linear power spectrum, $P(k)$ of `density'
perturbations comes from a number of distinct cosmological
observations of spectrum of cosmological perturbations . At low wave
numbers $k$, the $P(k)$ is inferred from the angular power spectrum of
the CMB anisotropy.  At the high $k$ end, the $P(k)$ is measured from
the Ly-$\alpha$ forest. The intermediate regime, is probed by
distribution of galaxies in the large scale surveys (here,SDSS). The
measurements from these different measurements line up along the
theoretical predictions of the currently favored cosmology. }
\label{fig:kspace}
\end{figure}

In its totality, quantifying and understanding the observed large
scale structures in the universe involves complex, non-linear aspects
of gravitational instability and baryon `gastrophysics'. However, in
estimating cosmological parameters the observational constraints from
structure formation on scales larger than $\sim$10$h^{-1} \ {\rm Mpc}$
are dominant.  At present, the constraints come from the measured
linear power spectrum of density perturbations $P(k)$. As shown is
figure 2, for small values of the wave number, $P(k)$ are probed by
the CMB anisotropy spectrum. On the intermediate wave numbers, $P(k)$
is measured by the ongoing large surveys of galaxies such as the SDSS
and 2 degree field~\cite{sdss,2df}. The power spectrum at the largest
wave numbers considered here is measured from the one-dimensional
distribution of absorption features along the line of sight to
quasars~\cite{lya}.

There is a well-understood (if not rigorously defined) notion of a
`standard' model of cosmology that includes the formation of large
scale structure.  Table 1 is an illustrative list of
parameters that characterize a cosmology in terms of background
evolution as well as structure formation~\cite{maxsdss}. Variations on
the choice, as well as, combinations of the parameters are possible
and have been used.  More importantly, these constitute a kind of
`minimal' accepted set. We often need to extend the set of parameters.
In particular, it is important to distinguish between the cosmological
parameters and the parameters characterizing the initial conditions
(IC) for primordial perturbations. The dimensionality of the IC sector
is largely kept under check by an implicit adherence to the generic
predictions of the simplest inflationary scenarios.

\begin{table}[h!]
\label{parlist}
\caption{ Cosmological parameters used in the parameter estimation
with WMAP and SDSS data using Markov chain Monte Carlo
technique(MCMC) [3]. The first 13 parameters
determine the rest of the parameters (below the dividing line). The
parameters labeled IC characterize the initial condition of the
primordial perturbations. }
\vspace*{0.5cm}

\begin{tabular}{lll}
&&\\[-2mm]
Parameter		&Meaning	&Definition\\
&&\\[-2mm]
\hline
&&\\[-2mm]
$\ob   		$&Baryon density			&$\ob=\Ob h^2$\\
$\od   		$&Dark matter density		&$\od=\Od h^2 $\\
$\fn	        $&Dark matter neutrino fraction	&$\fn=\rho_\nu/\rho_d$\\
$\Ol 		$&Dark energy density		&\\
$w	        $&Dark energy equation of state	& $p_\Lambda/\rho_\Lambda$ \\
                 & &(approximated as constant)\\
$\Ok       	$&Spatial curvature		&\\
$\tau 	        $&Reionization optical depth	&\\
$\As            $&Scalar fluctuation amplitude[IC]	&Primordial scalar power \\
&& at chosen pivot $k_*=0.05$/Mpc\\
$\ns	        $&Scalar spectral index	[IC]	&Primordial spectral index at $k*$\\
$\alpha         $&Running of spectral index[IC]	&$\alpha=d\ln\ns/d\ln k$ \\ 
&&(approximated as constant)\\
$r              $&Tensor-to-scalar ratio[IC]		&Tensor-to-scalar power ratio at $k_*$\\
$\nt           	$&Tensor spectral index	[IC]	&\\
$\zion        	$&Reionization redshift (abrupt)	&$\zion\approx 92 (0.03h\tau/\ob)^{2/3}\Om^{1/3}$ \\
&& (assuming abrupt reionization)\\
$\om    	$&Physical matter density	&$\om=\ob+\od = \Om h^2$\\
$\Om       	$&Matter density/critical density &$\Om=1-\Ol-\Ok$\\
$\Ot       	$&Total density/critical density  &$\Ot=\Om+\Ol=1-\Ok$\\
$\At	        $&Tensor fluctuation amplitude	&$\At=r\As$\\
$\Mnu    	$&Sum of neutrino masses 	&$\Mnu\approx(94.4\>{\rm eV})\times\od\fn$\\
$h	        $&Hubble parameter		&$h = \sqrt{(\od+\ob)/(1-\Ok-\Ol)}$\\
$t_0    	$	&Age of Universe & \\		
$\sigma_8       $&Galaxy fluctuation amplitude
&$\sigma_R^2=\int_0^\infty W[kR]^2 P(k) {k^2
dk\over2\pi^2}$, \\ 
& within sphere  $R= 8h^{-1}$Mpc & $W(x)=3(\sin x-x\cos x)/x^3$\\[1mm]
\end{tabular}
\end{table}

\begin{figure}
\vspace{-1.7cm}
\plotone{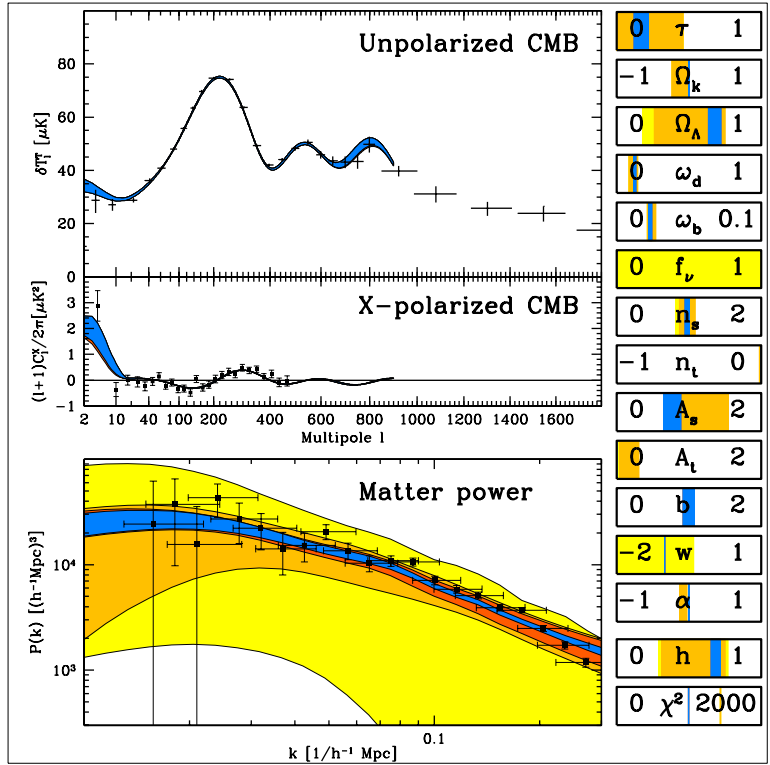}
\vspace*{0.5cm}

\caption{Summary of observations and cosmological models taken from
the combined analysis of WMAP and SDSS data~\protect\cite{maxsdss}.  Data
points are for unpolarized CMB experiments combined (top; Appendix A.3
details data used) cross-polarized CMB from WMAP (middle) and Galaxy
power from SDSS (bottom).  Shaded bands show the 1-sigma range of
theoretical models from the Monte-Carlo Markov chains, both for
cosmological parameters (right) and for the corresponding power
spectra (left).  From outside in, these bands correspond to 1.) WMAP
with no priors; 2.) adding the prior of zero contribution from
neutrinos $\fn=0$, and a cosmological constant $w=-1$, 3.)  adding the
priors $\Ok=0$ (flat universe), negligible gravity wave contribution
$r=0$ and power law primordial spectra $\alpha=0$; and finally adding
the SDSS information, respectively.  These four bands essentially
coincide in the top two panels, since the CMB constraints were
included in the fits.  Note that the $\l$-axis in the upper two panels
goes from logarithmic on the left to linear on the right, to show
important features at both ends, whereas the $k$-axis of the bottom
panel is simply logarithmic. }
\label{fig:omgl}
\end{figure}

Table 2 summarizes the current estimates of some of the
cosmological parameters based on the combined analysis of CMB
anisotropy measured by WMAP and the power spectrum of density
perturbations measured by SDSS and the Ly-$\alpha$
forest~~\cite{maxsdss,sel04}. The estimates given in the three columns
reflect the dependence of cosmological parameter estimates on the
choice and space of parameters. Figure~\ref{fig:omgl} summarizes the
observations and best fit cosmological models arrived at from the
combined analysis of WMAP and SDSS data.

Any observational comparison based on the structure formation in the
universe necessarily depends on the assumed initial conditions
describing the primordial seed perturbations of the perturbed
universe. What has been remarkable is the extent to which recent
cosmological observations have been consistent with and, in certain
cases, even vindicated the standard assumptions. Besides, the entirely
theoretical motivation of the paradigm of inflation, the assumption of
Gaussian, random adiabatic scalar perturbations with a nearly scale
invariant power is also arguably the simplest possible choice
for the initial perturbations.

\begin{table}[h]
\label{parval}
\caption{The current estimates of a selected set of cosmological
parameters obtained from the CMB anisotropy and LSS power spectrum
measurement from SDSS and Ly-$\alpha$\protect\cite{sel04}. The results
assume a flat universe and a cosmological constant (constant equation
of state $w=-1$). The second and third columns correspond to parameter
estimates when the space of IC parameters is curtailed by imposing the
assumptions of negligible gravity wave contribution $r=0$ and exact
power law primordial spectra $\alpha=0$.  The error bars quoted is the
larger of the upper and lower error bars ($1\sigma$ equivalent). For
upper bounds the $95$\% CL is quoted. See~\protect\cite{sel04} for
exact error bars at $1\sigma$, $2\sigma$ and $3\sigma$ equivalent
confidence levels. }
\vspace*{0.5cm}

\begin{tabular}{llll}
&&&\\[-2mm]
Parameter	& Priors ($\Omega_K=0$, $w=-1$)	
&+priors ($r=0$)  & +priors ( $\alpha=0$)  \\
&&&\\[-2mm]
\hline
&&&\\[-2mm]
$10^2 \ob$ & $2.42\pm  0.12$ & $2.33\pm 0.09$&$2.31 \pm 0.09$  \\
$\Omega_m$ & $0.27\pm 0.037$ &  $0.28\pm 0.022$& $0.3\pm 0.04$ \\
$h$        & $0.72\pm 0.037$ & $0.71\pm 0.02$ &$0.694 \pm 0.03$ \\
$\tau$     & $0.14\pm 0.049$ & $0.163\pm 0.04$&$0.133 \pm 0.05$ \\
$\sigma_8$ & $0.9 \pm 0.034$  & $0.9\pm 0.03$ &$0.89\pm 0.03$  \\
$n_s$      & $0.97\pm 0.023$  & $0.98\pm 0.02$& $1.00\pm 0.03$ \\
$10^2 \alpha$   & $-0.57\pm 1.2$ & $-0.3\pm 1.1$& fixed \\
$r$        &  $< 0.45$ & fixed  & fixed  \\
&&&\\[-2mm]
\end{tabular}
\end{table}

In a simple power law parametrization of the primordial spectrum of
density perturbation ($|\delta_k|^2 = A k^{n_s}$), the scale invariant
spectrum corresponds to $n_s=1$. Recent estimation of (smooth)
deviations from scale invariance favor a nearly scale
invariant spectrum~\cite{sel04}.  Current observations
favor a value of $n_s = 
0.98\pm0.02$ (99.9\%CL) which  are consistent with a
nearly scale invariant power spectrum. The current combined CMB and
LSS data is good enough to constrain the `running' of the spectral
index, $\alpha= {\rm d} n_s/{\rm d}\ln k = 0.003\pm 0.01$ 
(99.9\%CL). These
results are remarkably consistent with the generic predictions of the
simplest models of inflation.  The power in the CMB temperature
anisotropy at low multipoles ($l\lsim 60$) first measured by the
COBE-DMR~\cite{cobedmr} did indicate the existence of correlated
cosmological perturbations on super Hubble-radius scales at the epoch
of last scattering, except for the (rather unlikely) possibility of
all the power arising from the integrated Sachs--Wolfe effect along the
line of sight. Since the polarization anisotropy is generated only at
the last scattering surface, the negative trough in the
$C_l^{\rm TE}$
spectrum at $l\sim 140$ (that corresponds to a scale larger than the
horizon at the epoch of last scattering) measured by WMAP seals this
loophole, and provides an unambiguous proof of apparently `acausal'
correlations in the cosmological perturbations~\cite{ben_wmap03}.
Inflation is the most promising causal explanation for the generation
of these `acausal' correlations. Further, the negative power in
$C_l^{\rm TE}$ is a trademark of the adiabatic scalar metric perturbations
that is also a generic prediction of the simplest models of inflation.

Other than some anomalies reported regarding  the recent
WMAP results (see
below), the CMB anisotropy maps (including the WMAP non-gaussianity
analysis carried out by the WMAP team~\cite{kom_wmap03}) have been
found to be consistent with a statistically isotropic, Gaussian random
field. This assumption is theoretically motivated by
inflation~\cite{inflpert}.  The Gaussianity of the CMB anisotropy on
large angular scales directly implies Gaussian primordial
perturbations~\cite{initgauss}.

Finally, the most interesting and robust constraint obtained is that
on the spatial curvature of the universe. The combination of CMB
anisotropy, LSS and other observations can pin down the universe to be
spatially flat, $\Omega_{\rm K} =-0.02\pm 0.02$. These
results are further tightened 
when combined with the constraints from high red-shift supernova (SN
Ia). The connection between the geometry of space and the precise
location of acoustic peaks leads to the widespread belief that the CMB
data from WMAP  alone  can measure $\Omega_{\rm K}$ precisely. However, the
present CMB anisotropy from the WMAP spectrum alone constrains the
curvature rather weakly $-0.4 \lsim \Omega_{\rm K} \lsim 0.1$. This is due
to the degeneracy $\Omega_{\rm K}$ with the Hubble constant, $h$ and the age
of the universe, $t_0$. Even reasonably weak priors, $h > 0.5$ or
$t_0>12\,\,{\rm Gyr}$ tightens the constraint on
$\Omega_{\rm K}$
significantly.

The current constraints on curvature density eliminates one of the
three currently important players in eq.~(\ref{fried}) leading to
$\Omega_{\rm m} + \Omega_\Lambda \approx 1$. The physical origin and nature
of $\Omega_\Lambda$ is one of the major challenges of present
cosmology. The simplest parametrization of $\Omega_\Lambda$ as pure
vacuum energy density with an equation of state $w=-1$ is consistent
with all the observations~\cite{maxsdss}. This conclusion appears to
hold for the combination of CMB and LSS data in analyses where the
equation of state is allowed to vary with
red-shift~\cite{maxsdss,sel04}.

\section{Discussion}

Despite the precision of the cosmological parameter estimates, we do
not know them `accurately'. It is clear from Table 2 that there are
systematic differences in the median value depending on the choice of
priors and the space of parameters. There are significant differences
also due to the choice of the supplementary observational data sets
used.  The fact that the results differ when CMB anisotropy data are
combined with different `complementary' data sets also indicates that
the other observational constraints are not as reliable as the CMB
anisotropy~\cite{maxsdss}.

The dependence on the choice of the parameter space or the priors
imposed on them simply arises due to the covariances (and
degeneracies, therein) between parameters. Even imposing a uniform
prior but on different parameter combination can make a difference to
measurement of physical observables. For example, imposing uniform
prior on the epoch of reionization $z_{\rm ion}$, instead of the
optical depth $\tau$, tends to lower the estimate of
$\tau$~\cite{maxsdss,cont03}. Progress on new observational fronts
would help resolve these issues by breaking the parameter degeneracies
and eliminating/constraining extra parameters.  For example, it is
only very recently that the angular power spectrum of CMB polarization
has been detected. The degree angular scale interferometer (DASI) has
measured the CMB polarization spectrum over a limited band of angular
scales ($l\sim 200$--440) in late 2002~\cite{kov_dasi02}. The WMAP
mission has also detected CMB polarization~\cite{kog_wmap03}. WMAP is
expected to release the CMB polarization spectra very soon.  The
results of much awaited CMB polarization spectrum from WMAP would be a
crucial test of the early epoch of reionization and resolve some other
degeneracies. Future experiments that target the $B$-mode polarization
signature of gravity waves will be invaluable in pinning down the
values of $r$, and consequently, in identifying the viable sectors in the
space of inflationary parameters.

In principle, there is no well-defined procedure for selecting `the
set' of parameters. While there may be some consensus on the choice of
cosmological parameters, the situation becomes murky when confronted
with the parameters describing initial conditions (or, inflationary
parameters). In the absence of an accepted early universe scenario
(more narrowly, a favored model of inflation), it is difficult to {\it
a priori} set up and justify the chosen space of initial
conditions. The complex covariances between the cosmological and the
initial parameters are sensitive to the parametrization of the space
of initial spectra adopted.  Efforts along these lines are further
obscured by issues such as the applicability of the Occam's razor to
dissuade the extension of the parameter space of initial conditions. Such
deliberations have been recently framed in the more quantitative
language of Bayesian evidence to evaluate and select between possible
parametrizations~\cite{jaf03}.  However, this approach cannot really
point to a preferred parametrization. A possible approach may be to
maximize the likelihood over the space of initial conditions. This has
been suggested in the context of the primordial power
spectrum~\cite{arm_sour04}.  A similar situation exists regarding the
parametrization of the dark energy component.

The estimation of the parameters is only the first step. The challenge
that lies ahead would be to connect them to physics.  The physical
origin and properties of the dark energy component is a complete
mystery~\cite{coscons_rev}. While interpreting the cosmological
constant as vacuum energy may be the simplest parametrization, it
implies an extreme form of fine tuning.  Quintessence field does
alleviate the problem of fine tuning, but the scalar field does not
appear to have any other reason for its existence (in contrast to the
inflaton). So postulating a scalar field to explain one cosmological
observation appears to be a theoretical overkill.  The precise
property of the cold dark matter remains an open problem.  Eventually,
direct detection and identification of the CDM particle candidates is
perhaps needed. Meanwhile, cosmological observations pertaining to
successful galaxy formation are beginning to put interesting
constraints on the properties of the CDM~\cite{ost_stein03}. In
particular, simulations using canonical `collision-less' CDM appears
to be at odds with the observations on small (sub Mpc.) scales.
First, the substructure of CDM halos is predicted to be richer than
observed. The number of small galaxies that are observed orbiting with
a larger unit is less than expected. Second, the density profile at
the centers of CDM halos is predicted to be `cuspier' than
observed. Addressing these issues is complex, since predictions come
from large N-Body and hydrodynamic simulation which have their own
limitations. At this time, we are still mired in uncertainty.
Alternative variants to the collision-less CDM are under active
investigation. They include possibilities such as self-interacting
dark matter ~\cite{sidm}, warm dark matter ~\cite{wdm}, self
annihilating dark matter~\cite{sadm}, massive black holes~\cite{bhdm},
etc. There is an interesting phenomenology developing as sub-field of
cosmology.  It is not inconceivable that cosmological observations
would pin down the properties of the CDM component well before any
direct detection.

There could also be major surprises hidden in the current cosmological
observations.  After the recent release of the first year of WMAP
data, anomalies such as the suppression of power in the lowest
multipoles of the CMB anisotropy, possible breakdown of statistical
isotropy (SI) and Gaussianity has attracted much attention.
Tantalizing evidence for SI breakdown (albeit, in very different
guises) has mounted in the {WMAP} first year sky maps, using a variety
of different statistics. It was pointed out that the suppression of
power in the quadrupole and octopole are aligned~\cite{maxwmap}, but
this could be due to imperfect (galactic) foreground  subtraction.
Further `multipole-vector' directions associated with these multipoles
(and, some other low multipoles as well) appear to be anomalously
correlated~\cite{cop04,schw04}. There are indications of asymmetry in
the power spectrum at low multipoles in opposite
hemispheres~\cite{erik04a,han04}. Possibly related, are the results of
tests of Gaussianity that show asymmetry in the amplitude of the
measured genus amplitude (at about $2$ to $3\sigma$ significance)
between the north and south galactic
hemispheres~\cite{par04,erik04b}. Analysis of the distribution of
extrema in WMAP sky maps has indicated non-Gaussianity, and to some
extent, violation of statistical isotropy~\cite{lar_wan04}.  These
anomalies could be pointing to new physics with exciting cosmological
ramifications and need to be addressed with specialized statistical
measures~\cite{haj_sour}.

\section{Summary}

Observational cosmology has made impressive progress in recent years.
We now have the ability to make precise measurements of cosmological
parameters by including observational constraints from structure
formation in the universe.  While ability to make quantitatively
`precise' statements within a parameterized standard cosmology has
improved remarkably, we still have far to go before we are
`accurate' in describing our cosmos. Observational cosmology has to
first grapple with the physical interpretation of these precisely
measured parameters!

\end{document}